\documentstyle[epsfig]{elsart}
\begin{document}
\runauthor{STAR Collaboration }
\begin{frontmatter}
\title{Energy and system size dependence of $\phi$ meson production in Cu+Cu and Au+Au collisions}

 \author[uic]{B.I.~Abelev},
 \author[pu]{M.M.~Aggarwal},
 \author[vecc]{Z.~Ahammed},
 \author[kent]{B.D.~Anderson},
 \author[dubna]{D.~Arkhipkin},
 \author[jinr]{G.S.~Averichev},
 \author[nikhef]{Y.~Bai},
 \author[mit]{J.~Balewski},
 \author[uic]{O.~Barannikova},
 \author[uuk]{L.S.~Barnby},
% ...

(STAR Collaboration)

% ...
\address[uuk]{University of Birmingham, Birmingham, United Kingdom}
% ...
\address[uic]{University of Illinois at Chicago, Chicago, Illinois 60607}
% ...
\address[jinr]{Laboratory for High Energy (JINR), Dubna, Russia}
\address[dubna]{Particle Physics Laboratory (JINR), Dubna, Russia}
% ...
\address[kent]{Kent State University, Kent, Ohio 44242}
% ...
\address[mit]{Massachusetts Institute of Technology, Cambridge, MA 02139-4307}
% ...
\address[nikhef]{NIKHEF and Utrecht University, Amsterdam, The Netherlands}
% ...
\address[pu]{Panjab University, Chandigarh 160014, India}
% ...
\address[vecc]{Variable Energy Cyclotron Centre, Kolkata 700064, India}
% ...


\date{\today}
\begin{abstract}
We study the beam-energy and system-size dependence of $\phi$ meson production 
(using the hadronic decay mode $\phi$ $\rightarrow$ $K^{+}$$K^{-}$) by comparing 
the new results from Cu+Cu collisions and previously reported Au+Au collisions 
at  $\sqrt{s_{\mathrm {NN}}}$~=~62.4 and 200~GeV measured in the STAR experiment at RHIC. 
Data presented are from mid-rapidity ($\mid$$y$$\mid$~$<$0.5) for 0.4~$<$~$p_{\mathrm T}$~$<$~5~GeV/$c$.  
At a given beam energy, the transverse momentum distributions for $\phi$ mesons are observed 
to be similar in yield and shape for Cu+Cu and Au+Au colliding systems with similar average 
numbers of participating nucleons. The $\phi$ meson yields in nucleus-nucleus collisions, 
normalised by the average number of participating nucleons, are found to be enhanced relative 
to those from $p$+$p$ collisions with a different trend compared to strange baryons. The 
enhancement for $\phi$ mesons is observed to be higher at $\sqrt{s_{\mathrm {NN}}}$~=~200 GeV 
compared to 62.4 GeV. These observations for the produced $\phi (s\bar{s})$ mesons clearly suggest
that, at these collision energies, the source of enhancement of strange hadrons is related to the 
formation of a dense partonic medium in high energy nucleus-nucleus collisions and cannot be alone 
due to canonical suppression of their production in smaller systems.
\end{abstract}

\end{frontmatter}

\section{Introduction}

Experimental results from the Relativistic Heavy Ion Collider (RHIC) have
confirmed the formation of a hot and dense medium in the initial
stages of high-energy heavy-ion collisions~\cite{rhicwhitepapers}. 
Thus one of the prerequisites for the formation of a 
Quark Gluon Plasma (QGP)~\cite{shuryak} in such collisions has been established. 
High statistics data on $\phi$ meson elliptic flow and yields as a 
function of transverse momentum ($p_{\mathrm T}$) have been used to
support the picture of formation of a hot and dense medium with
partonic collectivity at RHIC~\cite{phiprl}. Evidence of $\phi$ mesons
being formed by the coalescence of seemingly thermalized $s\bar{s}$-quarks
in central Au+Au collisions has also been presented~\cite{phiprl}.

Several interesting features were also observed in the centrality 
dependence of $\phi$ meson production in Au+Au collisions at 200 GeV. 
As one goes from central collisions (average number of participants, $\langle N_{\mathrm {part}} \rangle$, $>$ 166) 
to peripheral collisions ($\langle N_{\mathrm {part}} \rangle$ $<$ 77), the $p_{T}$ spectra showed a gradual 
evolution from an exponential shape to a shape which requires an additional power law type of 
behavior at higher $p_{T}$ ($>$ 3 GeV/$c$)~\cite{phiprl,phiplb}.
At the same time, the average transverse momentum ($\langle$$p_{\mathrm T}$$\rangle$) of $\phi$ mesons, 
dominated by the transverse momentum distribution at low $p_{\mathrm T}$, showed no
significant collision centrality dependence in Au+Au collisions, unlike what has been seen for other 
particles of similar mass such as anti-protons ($\bar{p}$)~\cite{phiplb}. 
The $N(\phi)$/$N(K^{-})$ ratio was observed to be independent of collision centrality in
Au+Au collisions, in contrast to predictions from microscopic transport models like RQMD and UrQMD~\cite{urqmd}. 
Both of these results led to the conclusion that $\phi$ meson production may not be 
from $K\bar{K}$ coalescence and $\phi$ mesons may have decoupled early on in the collisions~\cite{phiplb}. 

The linear increase of the $N(\Omega)/N(\phi)$ ratio with $p_{\mathrm T}$ was proposed as an observable 
to test the recombination picture and hence also provided a test for thermalization in 
heavy-ion collisions~\cite{phireco}. A distinct trend was observed in the centrality dependence of this 
ratio vs. $p_{\mathrm T}$ in Au+Au collisions~\cite{phiprl}. With decreasing centrality, the observed $N(\Omega)/N(\phi)$
ratio seems to turn over at successively lower values of $p_{\mathrm T}$ indicating a smaller contribution
from thermal quark coalescence in more peripheral collisions.
Furthermore, in lower energy collisions at the SPS~\cite{na49}
and AGS~\cite{ags}, it was observed that the relative strangeness production increases with $N_{\mathrm {part}}$. 
For similar $N_{\mathrm {part}}$,
the increase was found to be slower for larger colliding ions. The possible reason was
related to variations of space-time density of the participating nucleons
and the increase in collision density (interactions per fm$^{3}$) towards the center of the reaction volume~\cite{na49,ags}.
The measurement of $\phi$ production in Cu+Cu collisions, 
in which systems with $N_{\mathrm {part}}$ $<$ 128 are created, is therefore expected to provide more precise data to further probe
these centrality and colliding ion size dependent features.

In this letter we report the first results of $\phi$ meson production 
for rapidities $\mid$$y$$\mid$~$<$~0.5 and  0.4~$<$~$p_{\mathrm T}$~$<$~5 GeV/$c$ in
Cu+Cu collisions at $\sqrt{s_{\mathrm {NN}}}$~=~62.4 and 200 GeV. The data were taken by 
the STAR experiment at RHIC ~\cite{starnim}. A detailed comparative study of the energy
and system size dependence of $\phi$ meson production ($p_{\mathrm T}$ spectra, rapidity 
density and  $\langle p_{\mathrm T} \rangle$) 
is carried out using both the Cu+Cu and Au+Au data. 

Several possible mechanisms of $\phi$ meson production in nucleus-nucleus 
collisions have been reported
in the literature~\cite{phireco,shor,dover,sorge}.
Some of these are supported by the experimental data~\cite{phiprl} which is not true with others~\cite{phiprl}.
In a QGP, thermal $s$ and $\bar{s}$ quarks can be produced by gluon-gluon 
interactions~\cite{shor}. These interactions could occur very rapidly and the 
$s$-quark abundance would equilibriate. During hadronisation,
the $s$ and $\bar{s}$ quarks from the plasma coalesce to form $\phi$ mesons. 
Production  by this process is not suppressed as per the OZI (Okubo-Zweig-Izuka) rule~\cite{ozi}. 
This, coupled with large abundances of strange quarks in the plasma, may lead to a dramatic 
increase in the production of $\phi$ mesons and other strange hadrons 
relative to non-QGP $p$+$p$ collisions~\cite{seqgp}.
Alternative ideas of canonical suppression of strangeness in small systems as a source
of strangeness enhancement in high energy heavy-ion collisions have been proposed for 
other strange hadrons (e.g $\Lambda$, $\Xi$ and $\Omega$)~\cite{canonical}.
The strangeness conservation laws require the production of an $\bar{s}$-quark
for each $s$-quark in the strong interaction. The main argument in such canonical models is that the energy and space time 
extensions in smaller systems may not be sufficiently large. This leads to a suppression of strange hadron production in small
collision systems.
These statistical models fit the data reasonably well~\cite{canonical_expt}. 
According to these models, strangeness enhancement in nucleus-nucleus 
collisions, relative to $p$+$p$ collisions, should increase with the strange quark content of the hadrons. This 
enhancement is predicted to decrease with increasing beam energy~\cite{canonical_prediction}. 
So far, discriminating between the two scenarios 
(strange hadron enhancement being due to dense partonic medium formed in heavy-ion collisions or due to canonical
supression of their production in $p$+$p$ collisions) using the available experimental data has been, to some extent,
 ambiguous.
Enhancement of $\phi (s\bar{s})$ production (zero net strangeness) in Cu+Cu and Au+Au 
relative to $p$+$p$ collisions would clearly indicate the formation of a dense partonic medium in these collisions. 
This would then rule out canonical suppression effects being the most likely cause for the observed 
enhancement in other strange hadrons~\cite{enhance} in high energy heavy-ion collisions.

\section{Experiment and analysis}
\begin{figure*}
\begin{center}
\includegraphics[scale=0.5]{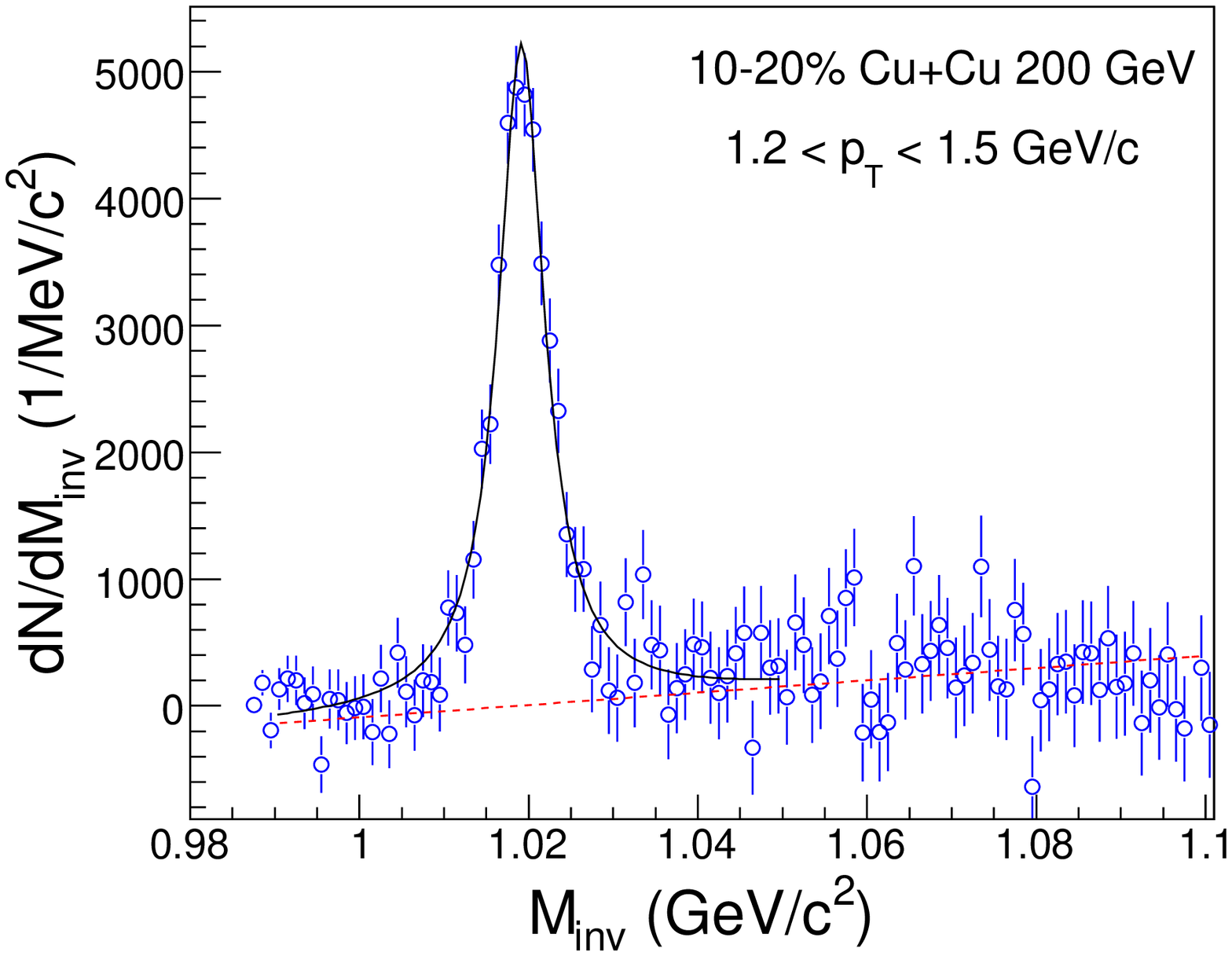}
\includegraphics[scale=0.5]{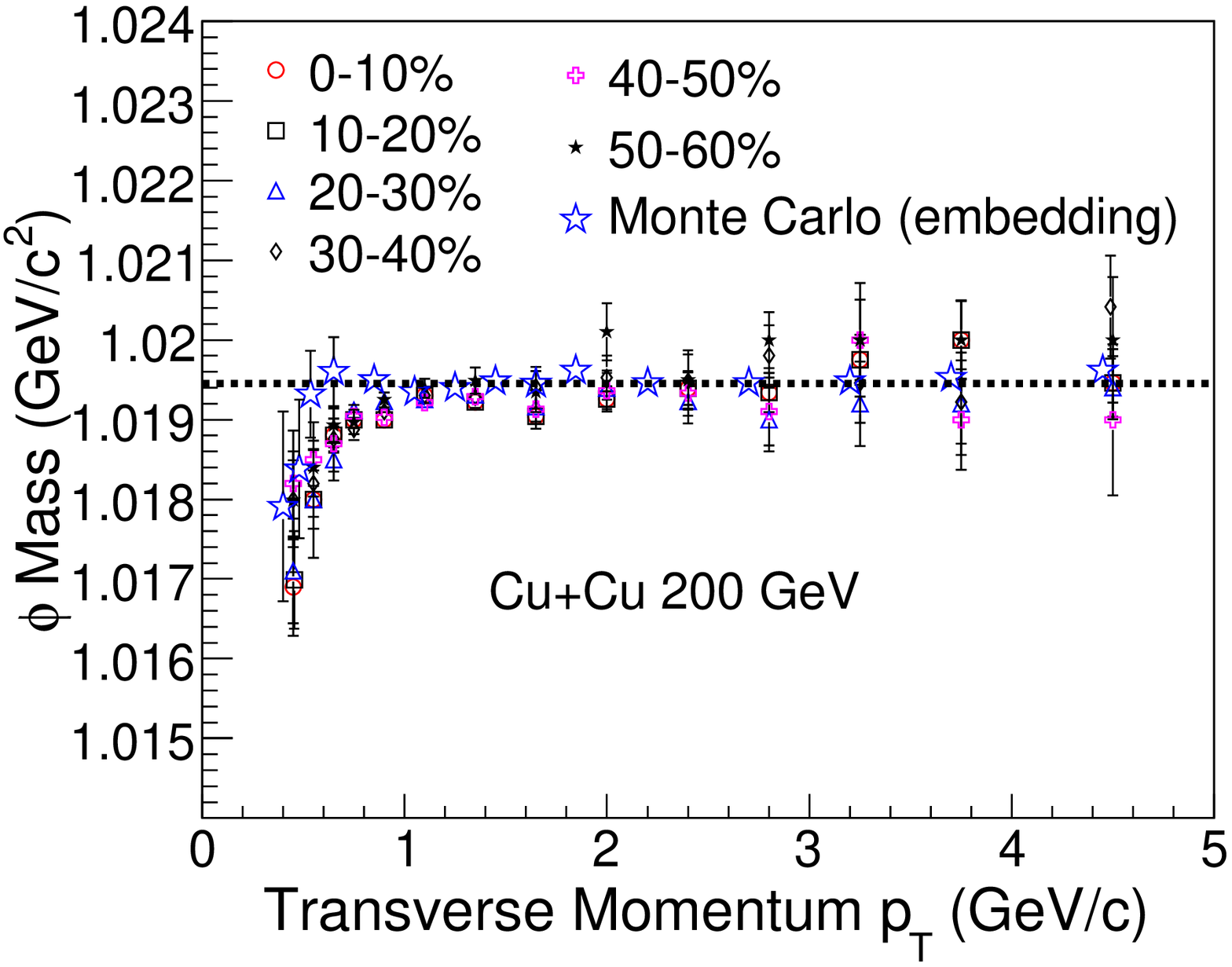}
\caption{Upper panel: A typical $\phi$ meson mass peak in Cu+Cu collisions at 200 GeV
obtained from the $K^{+}K^{-}$ invariant mass distribution after subtracting the
combinatorial background using mixed events. The distribution is fitted to a
Breit-Wigner function (solid line) and  a linear background function (dashed line)
to extract the yields. The errors shown are statistical. Lower panel: $\phi$ meson mass peak position
as a function  of $p_{\mathrm T}$ for various collision centralities in 
Cu+Cu collisions at 200 GeV. Also shown are the results from Monte Carlo calculations for 0-60\%
centrality 
using embedding techniques (see text for more details) shifted by 50 MeV/$c$ in $p_{\mathrm T}$ for clarity of presentation.
The dashed line corresponds to the PDG value of 1.0194 GeV/$c^{2}$~\cite{pdg}.}
\label{fig1}
\end{center}
\end{figure*}

\begin{table}
\caption{\label{table0}
Collision systems, beam energies, number of events and trigger conditions.}
\vspace{.5cm}
\begin{center}
\begin{tabular}{|c|c|c|c|c|}
\hline
Collision system & $\sqrt{s_{NN}}$ (GeV)  & Number of events &  Trigger condition \\
\hline
Cu+Cu  & 62.4   & 8.8 $\times 10^{6}$   & Minimum Bias \\
Cu+Cu  & 200    & 1.5 $\times 10^{7}$   & Minimum Bias \\
Au+Au  & 62.4   & 6.2 $\times 10^{6}$   & Minimum Bias \\
Au+Au  & 200    & 1.35 $\times 10^{7}$  & Minimum Bias \\
Au+Au  & 200    & 1.0 $\times 10^{7}$   & Central Trigger (0-12\%) \\
\hline
\end{tabular}
\end{center}
\end{table}

The data presented here were taken at RHIC in 2004 (Au+Au) and 2005 (Cu+Cu) 
using the STAR detector~\cite{starnim}. The analysis presented is from the data taken
by the Time Projection Chamber (TPC)~\cite{tpc}. The TPC magnetic 
field was 0.5 Tesla. Data were taken in both field configurations.
The trigger conditions and number of events analyzed for different colliding 
systems at $\sqrt{s_{\mathrm {NN}}}$~=~62.4 GeV and 200 GeV are given in Table~\ref{table0}.
The $\phi$ meson spectra for Au+Au collisions at 200 GeV using these data sets have been 
presented elsewhere.~\cite{phiprl}. Centrality selection for the Au+Au and Cu+Cu collisions 
utilized the uncorrected charged particle multiplicity for pseudorapidities $|\eta|$~$<$~0.5, measured 
by the TPC. Table~\ref{table1} shows the $\langle N_{\mathrm {part}} \rangle$ and $\langle N_{\mathrm {bin}}\rangle $ 
values calculated using a Glauber model
for different centralities for Cu+Cu collisions at 62.4 and 200 GeV and Au+Au collisions at 200 GeV.
The corresponding values for Au+Au collisions at 62.4 GeV were published previously~\cite{pmdftpc}.

The $\phi$ meson yield in each $p_{\mathrm T}$ bin was extracted from the invariant mass ($M_{inv}$)
distributions of $K^{+}K^{-}$ candidates after the subtraction of the combinatorial background
estimated using the event mixing technique~\cite{phiprl,phiplb,phdthesis}. The charged kaons were identified
through their ionization energy loss in the TPC. 
Figure~\ref{fig1} shows a typical, background subtracted, $K^{+}K^{-}$ $M_{inv}$ distribution as obtained for 200 GeV Cu+Cu
collisions. The 
resultant distribution is well described by a Breit-Wigner function (solid line) 
plus a linear background function (dashed line). The form of the Breit-Wigner function
is  $\frac{dN}{dM_{inv}} = \frac{C \Gamma}{(M_{inv} - m_{\phi})^{2} + \Gamma^{2}/4}$,
where $C$ is the area under the mass peak, $\Gamma$ is the full width at half maximum for the
distribution in GeV/$c^2$ and $m_{\phi}$ is the mass of the $\phi$ meson.
Figure~\ref{fig1} also shows that for $p_{\mathrm T}$ $>$ 0.7 GeV/$c$, the mass peak position of 
the $\phi$ meson agrees well with the PDG value of 1.0194 GeV~\cite{pdg}. For $p_{\mathrm T}$ $<$ 1.2 GeV/$c$ there
is a monotonic drop in the value of the fitted mass value with decreasing $p_{\mathrm T}$, reaching 
(mass $\phi$ fitted - mass $\phi$ PDG) $=$ -2.5 MeV at $p_{\mathrm T}$ $=$ 0.5 GeV/$c$.
The reconstructed invariant mass distribution of the $\phi$ meson is wider than the PDG value (4.26 MeV/$c^2$), 
decreasing from 9 MeV/$c^2$ to 4.26 MeV/$c^2$ with increasing  $p_{\mathrm T}$~\cite{philong}. 
The variations in the position of the
$\phi$ invariant mass peak and its width, at low $p_{\mathrm T}$,  are 
consistent with the simulation values and  are understood within the scope of the 
detector resolution effects~\cite{phdthesis}. To understand these effects,
$\phi$ decays to  $K^{+}$$K^{-}$ and 
detector response were studied within the STAR GEANT framework~\cite{stargeant}. The resulting simulated signals
were then embedded into real events before being processed by the standard STAR event 
reconstruction. These data were then processed like real data and analyzed to 
reconstruct the embedded $\phi$~\cite{phiprl,phiplb,phdthesis,philong}. 
Embedding simulations were also used to obtain
the $\phi$ meson  acceptance and reconstruction efficiency~\cite{phdthesis,philong}. The product of the acceptance 
and reconstruction efficiency was found to increase from
3\% at $p_{\mathrm T}$ = 0.5 GeV/$c$ to about 40\% at $p_{\mathrm T}$ = 3 GeV/$c$ for central Cu+Cu collisions.
The centrality dependence of these values were found to be  small for Cu+Cu collisions. At higher $p_{\mathrm T}$ (3-5 GeV/$c$), 
the efficiency was found to remain constant. 
The other important corrections applied to the data were related to the 
vertex finding efficiency which was $\sim$ 92.5\% and the correction for branching ratio 
of 49.2\% for the channel $\phi$ $\rightarrow$ $K^{+}K^{-}$.
A more detailed description of the $\phi$ meson mass peak position, width of the $\phi$ meson invariant mass 
distribution, variation of the reconstruction efficiency
with collision centrality and $p_{\mathrm T}$, and the general
procedure for obtaining the signal and constructing mixed events are discussed elsewhere~\cite{philong}.

\begin{table}
\caption{\label{table1}
The average numbers of participating nucleons ($\langle N_{\mathrm {part}} \rangle$) and binary collisions ($\langle N_{\mathrm {bin}\rangle }$)
for various collision centralities in Cu+Cu collisions at $\sqrt{s_{\mathrm {NN}}}$ = 62.4 and 200 GeV and Au+Au collisions at $\sqrt{s_{\mathrm {NN}}}$ = 200 GeV. }
\vspace{.5cm}
\begin{center}
\begin{tabular}{|c|c|c|c|c|c|c|}
\hline
\% cs & $\langle N_{\mathrm {part}}^{\mathrm {AuAu}} \rangle $ & $\langle  N_{\mathrm {bin}}^{\mathrm {AuAu}} \rangle$ & $\langle N_{\mathrm {part}}^{\mathrm {CuCu}} \rangle$ &  $\langle N_{\mathrm {bin}}^{\mathrm {CuCu}} \rangle$ & $\langle N_{\mathrm {part}}^{\mathrm {CuCu}} \rangle$ & $\langle N_{\mathrm {bin}}^{\mathrm {CuCu}} \rangle$ \\
      & 200 GeV           & 200 GeV          & 200 GeV           &  200 GeV          & 62.4 GeV          & 62.4 GeV         \\
\hline
0-10   & $325.9^{+5.5}_{-4.3}$  & $939.4^{+72.1}_{-63.7}$ & $99.0^{+1.5}_{-1.2}$ & $188.8^{+15.4}_{-13.4}$ & $96.4^{+1.1}_{-2.6}$ & $161.8^{+12.1}_{-7.5}$ \\
10-20  & $234.5^{+9.0}_{-7.8}$  & $590.9^{+60.8}_{-53.7}$ & $74.6^{+1.3}_{-1.0}$ & $123.6^{+9.4}_{-8.3}$ & $72.2^{+0.6}_{-1.9}$  & $107.5^{+6.3}_{-8.6}$ \\
20-30  & $166.6^{+10.1}_{-9.6}$ & $368.5^{+47.0}_{-44.9}$ & $53.7^{+1.0}_{-0.7}$ & $77.6^{+5.4}_{-4.7}$  & $51.8^{+0.5}_{-1.2}$ & $68.4^{+3.6}_{-4.7}$ \\
30-40  & $115.5^{+9.6}_{-9.6}$  & $220.1^{+35.1}_{-34.8}$ & $37.8^{+0.7}_{-0.5}$ & $47.7^{+2.8}_{-2.7}$  & $36.2^{+0.4}_{-0.8}$ & $42.3^{+2.0}_{-2.6}$ \\
40-50  & $76.7^{+9.0}_{-9.1}$   & $123.5^{+24.0}_{-25.4}$ & $26.2^{+0.5}_{-0.4}$ & $29.2^{+1.6}_{-1.4}$  & $24.9^{+0.4}_{-0.6}$ & $25.9^{+1.0}_{-1.5}$ \\
50-60  & $47.3^{+7.6}_{-8.1}$   & $63.9^{+15.5}_{-16.8}$  & $17.2^{+0.4}_{-0.2}$ & $16.8^{+0.9}_{-0.7}$  & $16.3^{+0.4}_{-0.3}$ & $15.1^{+0.6}_{-0.6}$ \\
60-70  & $26.9^{+5.5}_{-6.5}$   & $29.5^{+9.5}_{-9.8}$    &  --                 &  --                      & --                    & -- \\
70-80  & $14.1^{+3.6}_{-4.0}$   & $12.3^{+4.7}_{-4.8}$    &  --                 &  --                      & --                    & -- \\
\hline
\end{tabular}
\end{center}
\end{table}

Systematic errors for the $\phi$ meson spectral measurements in Cu+Cu collisions include uncertainties from the following sources: 
Uncertainties in $\phi$ meson reconstruction efficiency ($\sim$ 8-14\%), Kaon identification from $dE/dx$ (8\%),
Kaon energy loss corrections ($\sim$ 3-4\%), Residual background shape (4\%) and
magnetic field configuration ($\sim$ 3\%).
The systematic errors from all the above sources have been added in quadrature. Systematic errors for the 
$\phi$ meson spectra are similar at both energies (62.4 and 200 GeV). The total systematic errors for $\phi$ 
yields at both energies are estimated to be ${}^<_\sim$ 18\%
over the entire $p_{\mathrm T}$ range studied. A discussion on systematic
errors for Au+Au collisions, dN/dy, and $\langle$$p_{\mathrm T}$$\rangle$  can be found in Ref.~\cite{phiprl,phiplb,philong}.

\begin{figure*}
\begin{center}
\includegraphics[scale=0.7]{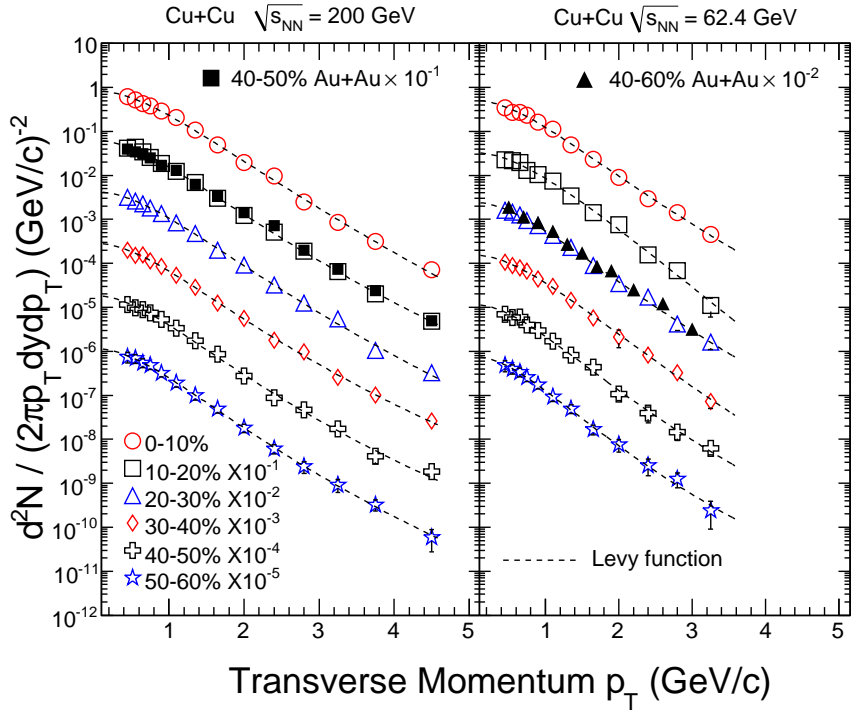}
\caption{Midrapidity ($\mid$$y$$\mid$~$<$~0.5) transverse momentum
spectra of $\phi$ mesons for various collision centrality classes for Cu+Cu  
collisions at $\sqrt{s_{\mathrm {NN}}}$~=~62.4  and 200 GeV. 
To study the system size dependence, comparison of 40-50\% Au+Au spectra 
to 10-20\% Cu+Cu spectra at 200 GeV, and 40-60\% Au+Au spectra to
20-30\% Cu+Cu spectra at 62.4 GeV are shown. These centralities for 
the two colliding systems have similar $\langle N_{\mathrm {part}} \rangle$ values as outlined in Table~\ref{table1}.
The errors represent the statistical and systematic errors added in 
quadrature. They are found to be within the symbol size. 
The spectra are fitted to a L\'{e}vy function discussed in the text.}
\label{fig2}
\end{center}
\end{figure*}

\section{Transverse momentum distributions and yields}

Figure~\ref{fig2} shows the $\phi$ meson yields from Cu+Cu collisions at 62.4 and 
200 GeV for 0.4 $<$ $p_{\mathrm T}$ $<$ 5~GeV/$c$ and  various collision 
centralities. The spectra are well described by a L\'{e}vy 
function of the form $\frac{d^{2}N}{2\pi p_{\mathrm T}dp_{\mathrm T}dy} = \frac{A}{[1 + (m_{T}-m_{\phi})/nT_{Levy}]^{n}}$,
where $m_{T}$ = $\sqrt{p_{\mathrm T}^{2}+m_{\phi}^{2}}$. $A$, $T_{\mathrm {Levy}}$, and $n$ are the parameters of the function.  
In the limiting case of 1/$n$ $\rightarrow$ 0, the L\'{e}vy distribution approaches an exponential function.
The parameters $T_{\rm Levy}$ and $n$ have similar values for the Cu+Cu and Au+Au
systems with similar $\langle N_{\mathrm {part}} \rangle$ at 200 GeV. This reflects the similar 
shape for the $\phi$ meson spectra in 
both collision systems at a given energy and $\langle N_{\mathrm {part}} \rangle$.
A comparison of $\phi$ mesons spectra for 40-50\% central Au+Au ($\langle N_{\mathrm {part}} \rangle$ = 76.7) 
and 10-20\% central Cu+Cu ($\langle N_{\mathrm {part}} \rangle$ = 74.6) collisions at 200 GeV is 
shown in Fig.~\ref{fig2} (left panel). Similar results for 40-60\% central Au+Au ($\langle N_{\mathrm {part}} \rangle$ = 59.9) 
and 20-30\% central Cu+Cu ($\langle N_{\mathrm {part}} \rangle$ = 51.8) collisions at 62.4 GeV are also shown in 
the same figure on the right panel.
The ratios of the $\phi$ meson $p_{\mathrm T}$ spectra for Au+Au and Cu+Cu systems 
with similar $\langle N_{\mathrm {part}} \rangle$ agree within $\sim$ 10\%. 
This is further quantified by studying their rapidity density ($dN/dy$)
and $\langle$$p_{\mathrm T}$$\rangle$ for both colliding systems. 

\begin{figure*}
\begin{center}
\includegraphics[scale=0.7]{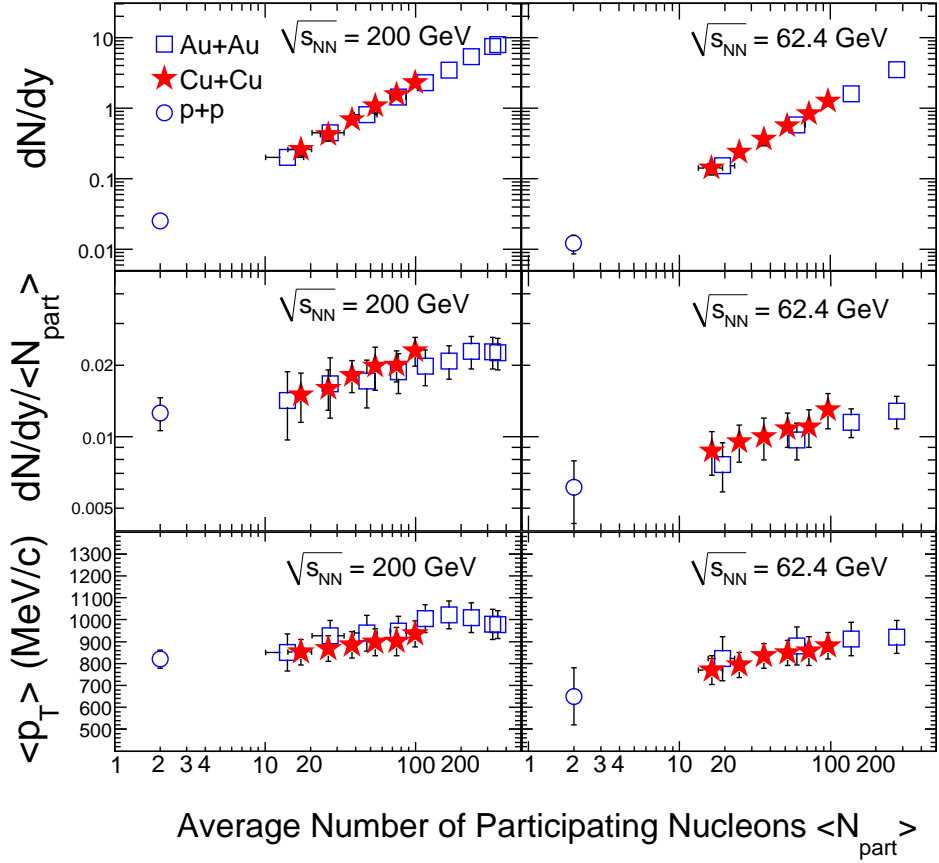}
\caption{Upper panels: $dN/dy$ at midrapidity for 
$\phi$ mesons for various collision centrality classes in Cu+Cu
and Au+Au at $\sqrt{s_{\mathrm {NN}}}$~=~200~GeV and 62.4 GeV. 
Also shown are the results from $p$+$p$ collisions.
Middle panels: same as above, but for $dN/dy/\langle N_{\mathrm {part}} \rangle$.
Lower panels: Average transverse momentum ($\langle$$p_{\mathrm T}$$\rangle$) for $\phi$ mesons at midrapidity 
for various event centrality classes for Cu+Cu  and Au+Au collisions 
at $\sqrt{s_{\mathrm {NN}}}$~=~62.4~GeV and 200 GeV. The $\langle$$p_{\mathrm T}$$\rangle$ for $\phi$ mesons
in $p$+$p$ collisions are also shown.
The error bars represent the statistical and systematic errors added in quadrature.}
\label{fig3}
\end{center}
\end{figure*}

Figure~\ref{fig3} shows $dN/dy$, $dN/dy/\langle N_{\mathrm {part}} \rangle$  and $<p_T>$ as a function of 
$\langle N_{\mathrm {part}} \rangle$ for Cu+Cu and Au+Au collisions at 62.4 and 200 GeV. 
Results from $p$+$p$ at 200 GeV and 63 GeV, obtained from the STAR~\cite{phiplb} 
and ISR ~\cite{isr} experiments respectively, are also included for comparison. 
At 63 GeV  the $d\sigma/dy$ for $\phi$ mesons 
at 0 $<$ $y$ $<$ 0.33 was reported to be 0.44 $\pm$ 0.11 (sys) $\pm$ 0.1 (stat) mb. 
These data, together with values of 36 and 42 mb for $p$+$p$ inelastic 
cross-sections at 63 and 200 GeV respectively, have been used to get
the corresponding $dN/dy$ values shown in the figure. 
The $dN/dy$ and $<p_T>$ values as obtained for the Cu+Cu collisions are 
also presented in Table~\ref{table2}.
Both at 62.4 and 200 GeV, all three quantities {\it viz} $dN/dy$, 
$dN/dy/\langle N_{\mathrm {part}} \rangle$ and $\langle$$p_{\mathrm T}$$\rangle$ 
scale with $\langle N_{\mathrm {part}} \rangle$.
These findings seem to indicate that the general features of $\phi$ meson production characterized
in terms of $dN/dy$ and $\langle$$p_{\mathrm T}$$\rangle$ at a given energy (62.4 or 200 GeV) 
do not depend on the colliding ion species studied, but depend on the $\langle N_{\mathrm {part}} \rangle$ of the collision. 
It will be interesting to see whether the same is true for other produced hadrons at RHIC. 
However, for a given $\langle N_{\mathrm {part}} \rangle$, both $dN/dy$ and $\langle$$p_{\mathrm T}$$\rangle$
 are observed to be lower for 62.4 GeV when compared to 200 GeV. 
This is in contrast to what has been seen at lower energies at AGS and SPS with smaller colliding systems~\cite{na49,ags}.
At those lower energies, for similar $\langle N_{\mathrm {part}} \rangle$, the strange hadron
production was higher while at RHIC, due to higher center of mass energy, a hotter and denser medium is expected to form
with a very low net baryon density at midrapidity~\cite{rhicwhitepapers}, leading to the observed differences.

\begin{table}
\caption{\label{table2}
$dN/dy$ and $\langle$$p_{\mathrm T}$$\rangle$  for $\phi$ mesons produced in Cu+Cu collisions 
at $\sqrt{s_{NN}}$ = 200 and 62.4 GeV for various collision centralities. The
errors include both systematic and statistical errors added in quadrature.}
\vspace{.5cm}
\begin{center}
\begin{tabular}{|c|c|c|c|c|}
\hline
\% centrality & $dN/dy$ & $dN/dy$   & $< p_{\mathrm T} >$ (MeV/$c$) &  $ <p_{\mathrm T} >$ (MeV/$c$) \\
      & 200 GeV & 62.4 GeV  & 200 GeV                   &  62.4 GeV            \\
\hline
0-10   & 2.3 $\pm$ 0.3   & 1.3 $\pm$ 0.2   & 935 $\pm$ 60 & 881 $\pm$ 61 \\
10-20  & 1.6 $\pm$ 0.2   & 0.8 $\pm$ 0.1   & 901 $\pm$ 64 & 857 $\pm$ 65 \\
20-30  & 1.1 $\pm$ 0.2   & 0.6 $\pm$ 0.1    & 897 $\pm$ 62 & 848 $\pm$ 57 \\
30-40  & 0.7 $\pm$ 0.1   & 0.4 $\pm$ 0.1    & 885 $\pm$ 60 & 835 $\pm$ 57 \\
40-50  & 0.4 $\pm$ 0.1   & 0.24 $\pm$ 0.04  & 869 $\pm$ 59 & 793 $\pm$ 57 \\
50-60  & 0.26 $\pm$ 0.05  & 0.14 $\pm$ 0.03 & 852 $\pm$ 58 & 771 $\pm$ 56 \\
\hline
\end{tabular}
\end{center}
\end{table}

\section{Nuclear modification factor}

Now we look at the $p_{\mathrm T}$ dependences of the nuclear modification factor, for the $\phi$ meson, 
both in terms of $N_{\mathrm {part}}$ and $N_{\mathrm {bin}}$. For $N_{\mathrm {part}}$, 
this factor is given by $$R_{\rm{AA}}^{N_{\rm part}}(p_{\rm T})\,=\,\frac{ d^2N_{\rm {AA}}/p_{\rm T}dy dp_{\rm T}/ \langle N_{\rm {part}}\rangle }{ d^2\sigma_{\rm {pp}}/dy dp_{\rm T}/\sigma_{\rm {pp}}^{inel} }.$$  To get the corresponding 
$R_{\rm{AA}}^{N_{\rm bin}}(p_{\rm T})$, one needs to replace $\langle N_{\rm {part}}\rangle$ by $\langle N_{\rm {bin}}\rangle$ 
in the above expression.
\begin{figure}
\begin{center}
\includegraphics[scale=0.6]{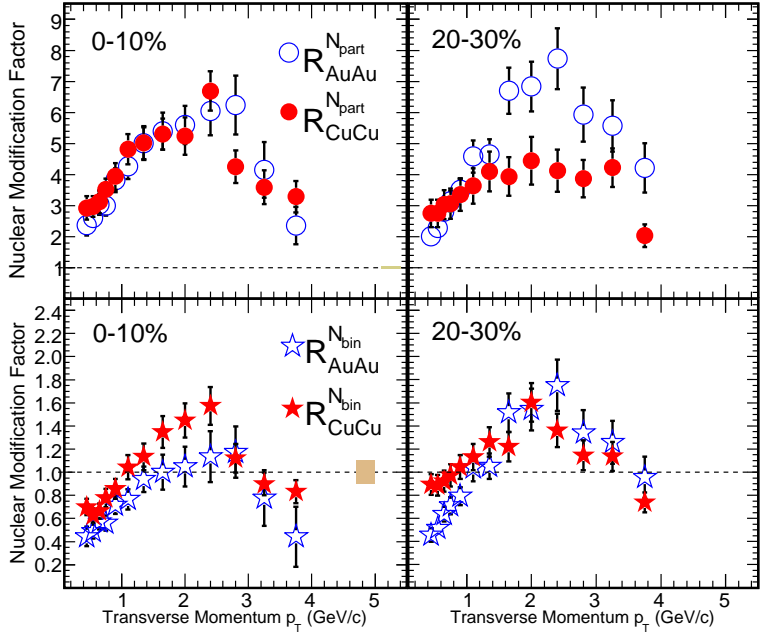}
\caption{Upper panels: $N_{\mathrm {part}}$ scaled ($R_{\mathrm {AA}}^{N_{\mathrm {part}}}$)
nuclear modification factor as a function of $p_{\mathrm T}$ of $\phi$ mesons for 0-10\% and
20-30\% Cu+Cu  and Au+Au collisions at $\sqrt{s_{\mathrm {NN}}}$ = 200 GeV. 
Lower panel: Same as above for $N_{\mathrm {bin}}$ scaled ($R_{\mathrm {AA}}^{N_{\mathrm {bin}}}$) nuclear modification factor.
The error bars represent the statistical and systematic errors added in quadrature.
The shaded band in upper panel around 1 at $p_{\mathrm T} = 4.5 - 5.5$ GeV/$c$ in the right side 
reflects the uncertainty in $\langle N_{\mathrm {part}} \rangle$  and that on the lower panel for 
$\langle N_{\mathrm {bin}}\rangle$ calculation for central Au+Au collisions. The respective 
uncertainties for central Cu+Cu collisions are of similar order.
}
\label{fig5}
\end{center}
\end{figure}
The results, as shown in Fig.~\ref{fig2} and Fig.~\ref{fig3} would lead to very similar results on 
$R_{\rm{AA}}^{N_{\rm {part}}}$ for both Cu+Cu and Au+Au systems for collisions having similar $\langle N_{\mathrm {part}} \rangle$.
In view of this, we only present here a comparison of the nuclear modification factors (in terms of $N_{bin}$ and $N_{part}$) 
of $R_{\mathrm {AA}}^{N_{\mathrm {bin}}}$ and $R_{\rm{AA}}^{N_{\rm {part}}}$ for Cu+Cu and Au+Au collisions.
For such a comparison only centralities corresponding to similar fraction of total hadronic cross-section 
were considered. 
The $R_{AA}$ for $\phi$ mesons  in 200 GeV Cu+Cu and Au+Au collisions for 
0-10\% and 20-30\% collision centralities (up to $p_{T}$ = 4 GeV/c) at 200 GeV are shown in Fig.~\ref{fig5}. 

Within the errors, the $R_{\rm{AA}}^{N_{\rm {part}}}$ values for 0-10\% central Cu+Cu and Au+Au collisions 
at 200 GeV
are seen to be similar in shape and yields. However, for 20-30\% collisions and at other collision centralities
(which are not shown in the figure) the Au+Au results are higher than Cu+Cu results for most of the $p_{\mathrm T}$ range studied.
The results for the central most Cu+Cu and Au+Au collisions studied are consistent with the observation 
that $dN/dy$/$\langle N_{\mathrm {part}} \rangle$  and $< p_{\mathrm T} >$ are constant as 
a function of $\langle N_{\mathrm {part}} \rangle$ for
$\langle N_{\mathrm {part}} \rangle$ $>$ 90 (Fig.~\ref{fig3}).

At the same collision centralities, the ratio $\langle N_{\mathrm {bin}}^{\mathrm {AuAu}} \rangle$/$\langle N_{\mathrm {bin}}^{\mathrm {CuCu}} \rangle$ is 
about $\sim$ 1.5 times larger than the ratio $\langle N_{\rm {part}}^{\mathrm {AuAu}} \rangle$/$\langle N_{\rm {part}}^{\mathrm {CuCu}} \rangle$.
This is reflected in the  $R_{\rm{AA}}^{N_{\rm bin}}$. As one can see from 
Fig.~\ref{fig5}, $R_{\rm{AA}}^{N_{\rm bin}}$ for 0-10\% Cu+Cu is higher than that of Au+Au collisions,
for  $p_{\mathrm T}$ $<$ 3 GeV/$c$.
Both the modification factors at $p_{\mathrm T}$ $>$ 3.5 GeV/$c$ are below unity, showing the characteristics of parton 
energy loss in hot and dense medium 
formed in central heavy-ion collisions. For 20-30\% central collisions, the similarity
 between $R_{\rm{AA}}^{N_{\rm bin}}$ for Cu+Cu and Au+Au collisions seems to extend to lower 
$p_{\mathrm T}$ ($\sim$ 1.5 GeV/$c$). 
It may be interesting to use the nuclear modification factor of $\phi$ mesons to investigate 
the differences in energy loss of quarks and gluons in the medium formed in 
heavy-ion collisions~\cite{colorch}. This is because $\phi$ mesons in central collisions are formed from 
coalescence of $s$ and $\bar{s}$ quarks~\cite{phiprl}, which presumably are formed by gluon interactions in
the initial stages of the collision.

\section{$\phi$ meson production and strangeness enhancement}
\begin{figure}
\begin{center}
\includegraphics[scale=0.7]{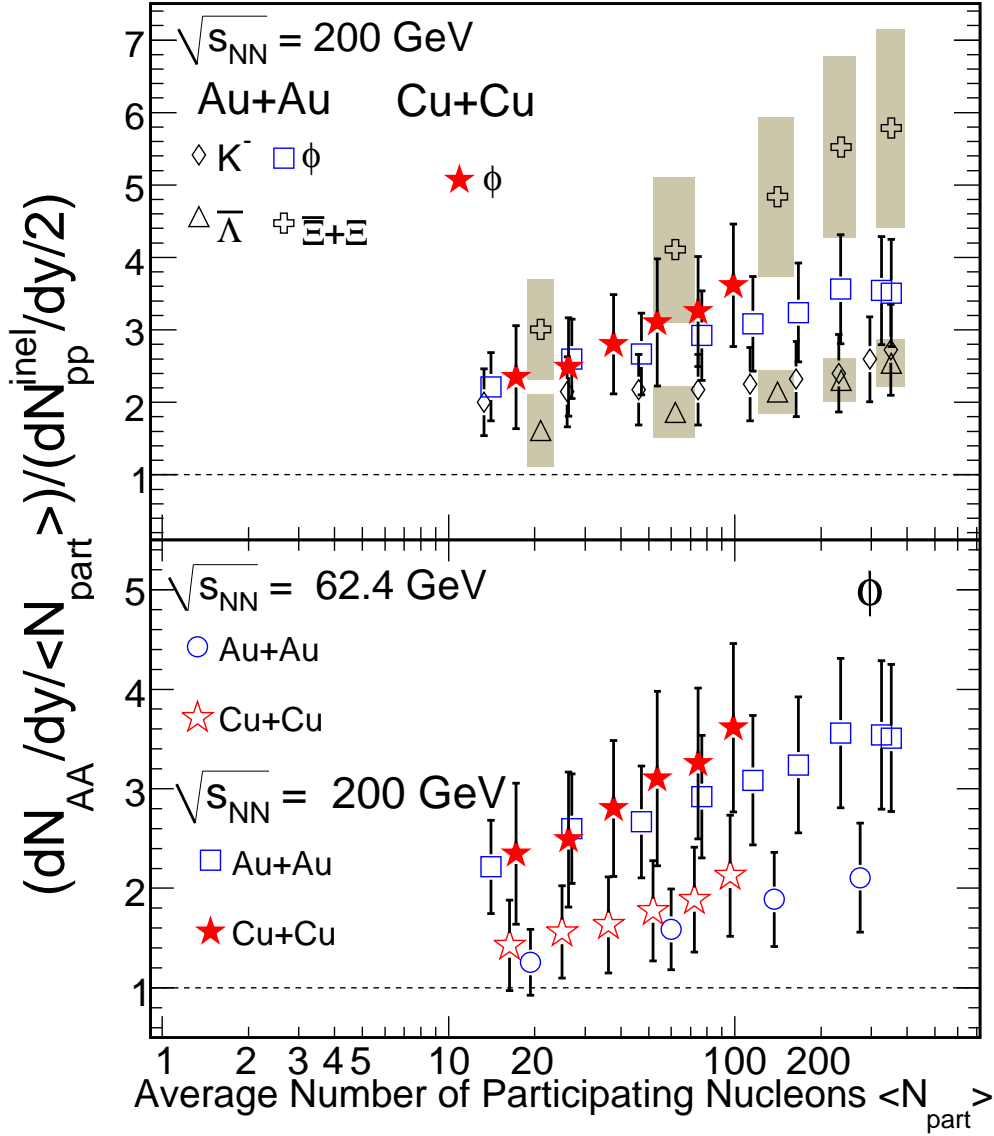}
\caption{Upper panel: The ratio of the yields of $K^{-}$, $\phi$, $\bar{\Lambda}$ and  $\Xi+\bar{\Xi}$
normalized to $\langle N_{\mathrm {part}} \rangle$ in nucleus-nucleus collisions to corresponding yields in inelastic $p$+$p$ 
collisions as a function of $\langle N_{\mathrm {part}} \rangle$ at 200 GeV.
Lower panel: Same as above for $\phi$ mesons in Cu+Cu collisions at 200 and 62.4 GeV.
The $p$+$p$ collision data at 200 GeV are from STAR~\cite{phiplb} and at 62.4 GeV from ISR~\cite{isr}. The error
bars shown represent the statistical and systematic errors added in quadrature.
}
\label{fig6}
\end{center}
\end{figure}

The ratio of strange hadron
production normalized to $\langle N_{\mathrm {part}} \rangle$ in nucleus-nucleus collisions relative to corresponding
results from $p$+$p$ collisions at 200 GeV are shown in the upper panel of Fig.~\ref{fig6}. The results are plotted as a function of $\langle N_{\mathrm {part}} \rangle$. 
$K^{-}$~\cite{pidprl}, $\bar{\Lambda}$ and  $\Xi+\bar{\Xi}$~\cite{enhance} are seen to 
show an enhancement (value $>$ 1) that increases with the number of strange valence quarks.
Furthermore, the observed enhancement in these open-strange hadrons increases with collision centrality, reaching
a maximum for the most central collisions. However,
the enhancement of $\phi$ meson production from Cu+Cu and Au+Au collisions shows a
deviation in ordering in terms of the number of strange constituent quarks. 
More explicitly, this enhancement is larger than for $K^{-}$ and $\bar{\Lambda}$, at the
same time being smaller than in case of $\Xi+\bar{\Xi}$.
Despite being different particle types (meson-baryon) and having different masses,
the results for $K^{-}$ and $\bar{\Lambda}$ are very similar in the entire centrality region studied. 
This rules out a baryon-meson effect as being the reason for the deviation of $\phi$ mesons 
from the number of strange quark ordering seen in Fig.~\ref{fig6} (upper panel).
The observed deviation is also not a mass effect as the enhancement in $\phi$ meson production 
is larger than that in $\bar{\Lambda}$ (which has mass close to that of the $\phi$).

In heavy-ion collisions, the production of $\phi$ mesons is not canonically suppressed 
due to its $s$$\bar{s}$ structure. In low energy $p$+$\bar{p}$ collisions at $\sqrt{s}$ = 3.6 GeV, 
$\phi$ meson production is suppressed due to the OZI rule~\cite{ppbar_ozi}. 
In $p$+$p$ collisions at $\sqrt{s}$ = 6.84 GeV violation of this rule has been
reported~\cite{pp_ozi}. At this higher energies $\phi$ production through 
channels accompanied by non-strange hadrons was found to dominate strongly over its production in channels 
accompanied with strange hadrons. Measurements of $\phi$ production in proton-nucleus collisions 
at $\sqrt{s_{NN}}$ = 27.4 GeV  have also shown that it takes place primarily by other 
than OZI allowed processes~\cite{ph_ozi}. Experiments studying inclusive $\phi$ production off protons 
by hadrons at incident momenta 63 and 93 GeV/$c$ also show that the production of $\phi$ mesons are
from OZI allowed processes~\cite{fusion_ozi}. Experiments on the production 
of $\phi$ mesons in $p$+$p$ collisions near threshold have shown a large enhancement of the
cross section ratio $\sigma(pp \rightarrow pp\phi)/\sigma(pp \rightarrow pp\omega)$ ~\cite{cosy} compared
to that predicted by the OZI rule~\cite{lipkin}. This ratio is sensitive to the basic feature of the
rule, which states that proceses with disconnected quark lines between initial and final states 
are suppressed compared to those where the incident quarks continue through to the exit channel.
The $p$+$p$ collisions at RHIC are at an energy which is $\sim$ 25 times higher than energies where 
violations of the OZI rule were reported~\cite{pp_ozi}. The $\phi$ meson enhancement in heavy-ion collisions 
shows an increasing trend with centrality (Fig.~\ref{fig6}).  From this, we conclude that 
the observed enhancement of  $\phi$ production in heavy-ion collisions may not be due to OZI 
suppression of $\phi$ production in $p$+$p$ collisions.

The observed enhancement of $\phi$ meson production then is a clear indication for the formation
of a dense partonic medium being responsible for the strangeness enhancement in Au+Au 
collisions at 200 GeV. Furthermore, $\phi$ mesons do not follow the strange quark ordering as 
expected in the canonical picture for the production of other 
strange hadrons. The observed enhancement in $\phi$ meson production being related to medium density 
is further supported by the energy dependence shown in the lower panel of  Fig.~\ref{fig6} . The $\phi$ meson 
production relative to $p$+$p$ collisions is larger at higher beam energy, a trend opposite to that 
predicted in canonical models for other strange hadrons. Earlier measurements have indicated that
$\phi$ meson production is not from coalescence of $K\bar{K}$ and minimally affected by 
re-scattering effects in the medium~\cite{phiplb}. Recent measurements indicate that 
$\phi$ mesons are formed from the coalescence of seemingly thermalized strange quarks~\cite{phiprl}.
All these observations put together along with the observed centrality and energy dependence of 
$\phi$ meson production (shown in Fig.~\ref{fig6}) indicate the formation of a dense partonic medium in 
heavy-ion collisions where strange quark production is enhanced (possible mechanisms could 
be as discussed in Refs.~\cite{shor,seqgp}). This in turn suggests that the observed centrality 
dependence of the enhancement for other strange hadrons (shown in Fig.~\ref{fig6}) is likely to be related to
the same reasons as in the case of the $\phi$ meson, that it is due to the formation of a dense partonic
medium in the collisions. These experimental data rule out the possibility of 
canonical suppression being the only source of the observed strangeness enhancement at beam energies of 200 GeV.

\section{Summary}

We have presented a study of the energy and system size dependence of 
$\phi$ meson production using the $p$+$p$, Cu+Cu and  Au+Au data at
$\sqrt{s_{\mathrm {NN}}}$ = 62.4 and 200 GeV. The $p_{\mathrm T}$
spectra are measured at midrapidity ($\mid$$y$$\mid$~$<$~0.5)
over the range  0.4~$<$~$p_{\mathrm T}$~$<$~5 GeV/$c$.
These measurements provide new experimental results showing that at a given 
beam energy the transverse momentum spectra in both shape ($\langle p_{\mathrm T} \rangle$) and
yields ($dN/dy$) are similar in Cu+Cu and Au+Au for collisions 
with similar $\langle N_{\mathrm {part}} \rangle$. In addition to observing similarity in the $\phi$ 
meson distributions for Cu+Cu and Au+Au collisions with similar $\langle N_{\mathrm {part}} \rangle$,
the $\langle N_{\mathrm {part}} \rangle $ scaled nuclear modification factors are observed to 
be similar for the 0-10\% central Cu+Cu and  Au+Au collisions at 200 GeV. However, 
such a similarity is not seen for other collision centralities. 
The corresponding results for the nuclear modification factor, scaled by the number of binary 
collisions, are in general found to be higher for Cu+Cu compared to Au+Au collisions.

The enhancement in the $\phi$ meson production has been studied through the ratio of the yields
normalized to $\langle N_{\mathrm {part}} \rangle$ in nucleus-nucleus collisions 
to corresponding yields in $p$+$p$ collisions as a function 
of $\langle N_{\mathrm {part}} \rangle$. The centrality and energy dependence of the enhancement in $\phi$ meson 
production clearly reflects the enhanced production of $s$-quarks in
a dense medium formed in high energy heavy-ion collisions. This then indicates that the
observed enhancements in other strange hadron ($K^{-}$,$\bar{\Lambda}$ 
and $\Xi+\bar{\Xi}$) production in the same collision system are likely to be due to
the similar effects and not only due to canonical suppression of strangeness production.
At RHIC the colliding beam energy is high, so it is very unlikely that the observed enhancement 
in heavy-ion collisions is due to OZI suppression of $\phi$ production in $p$+$p$ collisions.

The enhancement in the $\phi$ meson production deviates from the number of 
valence $s$-quark dependence observed for other strange hadrons.
The results from $\phi$ mesons lie in between those from single valence $s$-quark 
carrying hadrons $K^{-}$ and $\bar{\Lambda}$,
and double valence $s$-quark carrying hadrons $\Xi+\bar{\Xi}$. Comparisons with
other strange hadrons rule out the possibility of this being a baryon-meson or mass effect.
The exact reason for the observed deviation of the enhancement factor for the $\phi$ meson from the valence
strange quark dependence observed for other strange hadrons is not clear. It could be due 
to the effect of light-flavor valence 
quarks in the other strange hadrons or due to the net strangeness being zero in $\phi$ mesons.

We thank the RHIC Operations Group and RCF at BNL, and the NERSC Center 
at LBNL and the resources provided by the Open Science Grid consortium 
for their support. This work was supported in part by the Offices of NP 
and HEP within the U.S. DOE Office of Science, the U.S. NSF, the Sloan 
Foundation, the DFG Excellence Cluster EXC153 of Germany, CNRS/IN2P3, 
RA, RPL, and EMN of France, STFC and EPSRC of the United Kingdom, FAPESP 
of Brazil, the Russian Ministry of Sci. and Tech., the NNSFC, CAS, MoST, 
and MoE of China, IRP and GA of the Czech Republic, FOM of the 
Netherlands, DAE, DST, and CSIR of the Government of India, Swiss NSF, 
the Polish State Committee for Scientific Research,  and the Korea Sci. 
\& Eng. Foundation.

\end{document}